\documentclass[a4paper,11pt]{article}
\usepackage{pos}

\usepackage{bbold}
\usepackage{slashed}

\usepackage{acronym}

\newcommand{\abbrev}[1]{#1}

\newcommand{\myacrodef}[3]{\acrodef{#2}{#3}\newcommand{#1}{\ac{#2}}}

\myacrodef{\ckm}{CKM}{Cabbibo-Kobayashi-Maskawa}
\myacrodef{\gff}{GFF}{gradient flow formalism}
\myacrodef{\nlo}{NLO}{next-to-leading order}
\myacrodef{\nnlo}{NNLO}{next-to-next-to-leading order}
\myacrodef{\ope}{OPE}{operator product expansion}
\myacrodef{\sm}{SM}{Standard Model}
\newcommand{\qcd}{\abbrev{QCD}}

\usepackage[capitalize]{cleveref}

\newcommand{\citere}[1]{Ref.~\cite{#1}}
\newcommand{\citeres}[1]{Refs.~\cite{#1}}

\newcommand{\alphas}{\alpha_\mathrm{s}}
\newcommand{\EulerGamma}{\gamma_\text{E}}

\newcommand{\calo}{\mathcal{O}}
\newcommand{\tcalo}{\tilde{\calo}}
\newcommand{\Zchiring}{\mathring{Z}_\chi}

\newcommand{\ren}{\mathrm{R}}

\newcommand{\ctr}{T_\text{R}}
\newcommand{\tr}{\ctr}
\newcommand{\nc}{n_\text{c}}

\newcommand{\nf}{n_\text{f}}

\newcommand{\ep}{\epsilon}

\newcounter{notecount}

\title{The electroweak Hamiltonian in the gradient flow formalism}

\author[a]{Robert Harlander}
\author*[b,c]{Fabian Lange}

\affiliation[a]{TTK, RWTH Aachen University,\\
52056 Aachen, Germany}

\affiliation[b]{Institut f\"ur Theoretische Teilchenphysik, Karlsruhe Institute of Technology (KIT),\\
Wolfgang-Gaede Straße 1, 76128 Karlsruhe, Germany}

\affiliation[c]{Institut f\"ur Astroteilchenphysik, Karlsruhe Institute of Technology (KIT),\\
Hermann-von-Helmholtz-Platz 1, 76344 Eggenstein-Leopoldshafen, Germany}

\emailAdd{harlander@physik.rwth-aachen.de}
\emailAdd{fabian.lange@kit.edu}

\abstract{Over the last decade the gradient flow formalism has become an
  important tool for lattice simulations of Quantum Chromodynamics.  It
  offers remarkable renormalization properties which pave the way for
  cross-fertilization between perturbative and lattice calculations.  In
  this contribution we report on the construction of the flowed operator
  product expansion for the current-current operators of the electroweak
  Hamiltonian at \abbrev{NNLO} \qcd.  This allows for simpler transformations
  between lattice and perturbative schemes and might reduce the
  uncertainties of theoretical predictions for low-energy flavor
  observables.}

\FullConference{%
  *** The European Physical Society Conference on High Energy Physics (EPS-HEP2021), ***\\
  *** 26-30 July 2021 ***\\
  *** Online conference, jointly organized by Universität Hamburg and the research center DESY ***
}

\begin{document}
\maketitle

\section{Introduction}

Flavor observables play an important role in the search for new physics
beyond the \sm{}.
Theoretical predictions of such flavor observables are usually calculated
in the effective theory of weak interactions described by an effective
Hamiltonian~\cite{Buchalla:1995vs}.
Typically, the Wilson coefficients are derived from a perturbative
matching calculation to the \sm{} (or a particular model for
new physics). Depending on the process, their \sm{} expressions are known through \nlo{} or \nnlo{}, see e.g.~\citeres{Buchalla:1995vs,Bobeth:1999mk,Gorbahn:2004my,Brod:2010mj,Brod:2011ty}.
On the other hand, the evaluation of matrix elements of the
higher-dimensional composite operators
requires non-perturbative treatment as provided by lattice \qcd{}, for
example. Matching the renormalization conditions of these two factors is
an important source of uncertainties for theoretical predictions
in flavor physics~\cite{FlavourLatticeAveragingGroup:2019iem}.

The \gff{}~\cite{Luscher:2010iy} provides an elegant solution to this problem. It can be viewed
as a regularization scheme which applies to lattice as well as
perturbative calculations. Composite operators are finite after
renormalization of the \qcd\ parameters and the involved fields, without
the need for additional operator renormalization~\cite{Luscher:2011bx}. This implies that
there is no mixing among these operators under the renormalization
group.

In this contribution we consider the current-current operators.
Through a suitable basis transformation our results also apply to $|\Delta F|=2$ transitions such as $B\bar B$ or $K\bar K$ mixing.
We introduce the relevant
operators in \cref{sec:basis} and their flowed counterparts in
\cref{sec:flowed}.
In \cref{sec:EW-flope} we then derive the flowed \ope{}~\cite{Suzuki:2013gza,Makino:2014taa,Monahan:2015lha} for the electroweak Hamiltonian through \nnlo{}.
We conclude in \cref{sec:conclusions}.

Through \nlo{} this strategy has been performed in \citere{Suzuki:2020zue}, albeit in a different operator basis.

\section{Operator basis}\label{sec:basis}

We schematically write the effective Hamiltonian as
\begin{equation}
  \label{eq:EW-Hamiltonian} \mathcal{H}_\mathrm{eff} =
  - \frac{4G_\mathrm{F}}{\sqrt{2}} V_\mathrm{CKM} \sum_n
  C_n \calo_n
\end{equation}
where $G_\mathrm{F}$ denotes the Fermi constant and $V_\mathrm{CKM}$ the
relevant elements of the \ckm\ matrix~\cite{Buchalla:1995vs}.  The
Wilson coefficients $C_n$ capture the perturbative effects, whereas
matrix elements of the operators $\calo_n$ describe mostly
non-perturbative effects.

For the physical current-current operator basis we
choose~\cite{Chetyrkin:1997gb}
\begin{equation}
  \label{eq:EW_physical_operators} \begin{split}
  \calo_1 &=
  - \left(\bar\psi_{1,\mathrm{L}} \gamma_\mu
  T^a \psi_{2,\mathrm{L}}\right) \left(\bar\psi_{3,\mathrm{L}} \gamma_\mu
  T^a \psi_{4,\mathrm{L}}\right) ,\\ \calo_2
  &= \left(\bar\psi_{1,\mathrm{L}} \gamma_\mu \psi_{2,\mathrm{L}}\right)
  \left(\bar\psi_{3,\mathrm{L}} \gamma_\mu \psi_{4,\mathrm{L}}\right)
  , \end{split}
\end{equation}
where our convention for the color generators is\footnote{This
convention differs from the one of \citere{Chetyrkin:1997gb} which is
the reason for the sign difference in $\calo_1$ between that paper and
\cref{eq:EW_physical_operators}.}
\begin{equation}\label{eq:easy}
  \begin{split}
    [T^a,T^b] = f^{abc}T^c ,\qquad \text{Tr}(T^aT^b) = -\tr\delta^{ab}
  \end{split}
\end{equation}
and we adopt Euclidean space-time.  The subscript $\mathrm{L}$ denotes
the left-handed component of the spinors
\begin{equation}
  \psi_{\mathrm{R}/\mathrm{L}} = P_\pm \psi = \tfrac{1}{2} (1 \pm \gamma_5) \psi .
\end{equation}
The dependence of the fields on the space-time variable $x$ is
suppressed here and in what follows.  Throughout the calculation we
assume the quark flavors $\psi_1,\ldots,\psi_4$ to be mutually
distinct. All penguin operators as well as the chromomagnetic operator
do not contribute in this case.  By a suitable basis transformation, our
results apply also to the cases corresponding to $|\Delta
F|=2$ processes, i.e.\ $\psi_1 = \psi_3$ and/or $\psi_2 = \psi_4$ as
long as $\psi_2 \neq \psi_1 \neq \psi_4 \neq \psi_3\neq \psi_2$.

Working in dimensional regularization with $D=4-2\ep$, one has to
introduce so-called evanescent operators which vanish for $D = 4$, but
mix with the physical operators at higher orders in perturbation
theory~\cite{Buras:1989xd}.  However, one can prevent the physical
operators from mixing into the evanescent operators by a finite
renormalization such that all Green's functions of evanescent operators
vanish~\cite{Buras:1989xd,Dugan:1990df,Herrlich:1994kh}.  Our choice for
the evanescent operators is the same as in \citere{Chetyrkin:1997gb}.

\section{Flowed operators}\label{sec:flowed}

In the \gff, one defines flowed gluon and quark fields
$B^a_\mu=B^a_\mu(t)$ and $\chi=\chi(t)$ as solutions of the
flow-equations~\cite{Luscher:2010iy,Luscher:2013cpa}
\begin{equation}
  \label{eq:flow}
  \begin{gathered}
    \partial_t B^a_\mu = \mathcal{D}^{ab}_\nu G^b_{\nu\mu} + \kappa
    \mathcal{D}^{ab}_\mu \partial_\nu B^b_\nu ,\\ \partial_t \chi
    = \Delta \chi - \kappa \partial_\mu B^a_\mu T^a \chi ,\qquad \partial_t
    \bar \chi = \bar \chi \overleftarrow \Delta + \kappa \bar
    \chi \partial_\mu B^a_\mu T^a ,
  \end{gathered}
\end{equation}
with the initial conditions
\begin{equation}
  \begin{split}
    B^a_\mu (t=0) = A^a_\mu ,\qquad \chi (t=0)= \psi ,
    \label{eq:bound}
  \end{split}
\end{equation}
where $A^a_\mu$ and $\psi$ are the regular gluon and quark fields,
respectively, and
\begin{equation}
  \begin{gathered}
    \mathcal{D}^{ab}_\mu = \delta^{ab}\partial_\mu - f^{abc}
    B_\mu^c ,\qquad \Delta = (\partial_\mu + B^a_\mu T^a) (\partial_\mu
    + B^b_\mu T^b) ,\\ G_{\mu\nu}^a = \partial_\mu B_\nu^a -
    \partial_\nu B_\mu^a + f^{abc}B_\mu^bB_\nu^c .
  \end{gathered}
\end{equation}
The parameter $\kappa$ is arbitrary and drops out of physical
quantities; we will set $\kappa=1$ in our calculation, because this
choice reduces the size of the intermediate algebraic expressions.

Our practical implementation of the \gff\ in perturbation theory follows
the strategy developed in \citere{Luscher:2011bx} and further detailed
in \citere{Artz:2019bpr}. On the one hand, it amounts to generalizing
the regular \qcd\ Feynman rules by adding flow-time dependent
exponentials to the propagators. The flow equations, \cref{eq:flow}, are
taken into account with the help of Lagrange multiplier fields which are
represented by so-called \emph{flow lines} in the Feynman diagrams. They
couple to the (flowed) quark and gluon fields at flowed vertices which
involve integrations over flow-time parameters.

Replacing the regular by flowed quark fields in
\cref{eq:EW_physical_operators}, one arrives at the flowed physical operators
\begin{equation}
  \label{eq:EW_flowed_physical_operators}
  \begin{split}
    \tcalo_1 &= - \Zchiring^2\left(\bar\chi_{1,\mathrm{L}} \gamma_\mu
    T^a \chi_{2,\mathrm{L}}\right) \left(\bar\chi_{3,\mathrm{L}}
    \gamma_\mu T^a \chi_{4,\mathrm{L}}\right) ,\\ \tcalo_2 &=
    \Zchiring^2 \left(\bar\chi_{1,\mathrm{L}} \gamma_\mu \chi_{2,\mathrm{L}}\right)
    \left(\bar\chi_{3,\mathrm{L}} \gamma_\mu \chi_{4,\mathrm{L}}\right) ,
  \end{split}
\end{equation}
and analogously for the evanescent operators.
The
non-minimal renormalization constant $\Zchiring$ for the flowed quark
fields $\chi$ is defined by
\begin{equation}
  \label{eq:Zchiring}
  \begin{split}
    \mathring{Z}_\chi &= -\frac{2\nc \nf}{(4\pi t)^2} \cdot \frac{1}{\left.\langle R(t)\rangle\right|_{m=0}} , \\
    R(t) &= \sum_{f=1}^{\nf}\bar{\chi}_f(t) \overleftrightarrow{\slashed{\mathcal{D}}} \chi_f(t) , \qquad \overleftrightarrow{\mathcal{D}}_\mu = \partial_\mu
    -\overleftarrow{\partial}_\mu + 2B_\mu^a T^a ,
  \end{split}
\end{equation}
where $\langle\cdot\rangle$ denotes the vacuum expectation
value~\cite{Makino:2014taa}, $\nc$ is the number of colors, and $\nf$
the number of quark flavors.  $\mathring{Z}_\chi$ was computed through
second order in the strong coupling constant $\alphas$ in
\citere{Artz:2019bpr}.

\section{Flowed operator product expansion}
\label{sec:EW-flope}

The flowed operators can be expressed by the
\emph{small-flow-time expansion}~\cite{Luscher:2011bx}
\begin{equation}
  \label{eq:small-flow-time-expansion}
  \tcalo_i(t) \asymp \sum_j \zeta^\ren_{ij}(t) \calo^\ren_j ,
\end{equation}
where the symbol $\asymp$ signals that this relation holds
asymptotically in the limit $t\to 0$. The mixing matrix
$\zeta^\ren_{ij}(t)$ and the regular operators $\calo^\ren_j$ are
already renormalized here.  At the bare level, the mixing from and to
the evanescent operators has to be taken into
account~\cite{Buras:1989xd,Dugan:1990df,Herrlich:1994kh}.  Only after
renormalizing the regular operators can we discard the evanescent
operators in \cref{eq:small-flow-time-expansion}.

By inverting \cref{eq:small-flow-time-expansion} one finds the flowed \ope{}~\cite{Suzuki:2013gza,Makino:2014taa,Monahan:2015lha}.
In the case of the electroweak Hamiltonian of \cref{eq:EW-Hamiltonian} it reads
\begin{equation}
  \label{eq:EW-Hamiltonian-flowed}
  \mathcal{H}_\mathrm{eff} \asymp- \frac{4G_\mathrm{F}}{\sqrt{2}} V_\mathrm{CKM} \sum_n \tilde{C}_n \tcalo_n ,
\end{equation}
where the flowed Wilson coefficients are given through
\begin{equation}
  \label{eq:flowed-Wilson}
  \tilde{C}_n = \sum_m C^\ren_m (\zeta^{\ren})^{-1}_{mn} .
\end{equation}
Since the flowed operators $\tcalo_n$ do not require
renormalization~\cite{Luscher:2011bx}, the r.h.s.\ is scheme independent
and one can directly combine the perturbative results for the Wilson
coefficients with the matrix elements of the operators obtained by other
means, for example a lattice calculation.

By constructing suitable
\textit{projectors}~\cite{Gorishnii:1983su,Gorishnii:1986gn} (see also
\citere{Harlander:2018zpi}), we extracted the elements of the mixing
matrix $\zeta^\ren_{ij}(t)$ from \cref{eq:small-flow-time-expansion}.
Through \nnlo{} we find
\begin{equation}
  (\zeta^{\ren})^{-1} = \mathbb{1} +
  \frac{\alphas}{4\pi}\begin{pmatrix}
   16.85 & - 3.333\\
    - 15 & 14.85
  \end{pmatrix}
  + \left(\frac{\alphas}{4\pi}\right)^2
  \begin{pmatrix}
    363.5 - 11.55 \nf  & -72.49 + 2.521 \nf \\
    -371.2 + 11.35 \nf & 311.5 - 6.934 \nf
  \end{pmatrix}\,,
\end{equation}
where the \qcd\ color factors have been inserted and the
renormalization scale has been set to $\mu =
\frac{e^{-\EulerGamma/2}}{\sqrt{2t}}$.  We obtained these results in
general $R_\xi$ gauge and with the operator renormalization from the
literature, see e.g.\ \citere{Gambino:2003zm}.  Both facts provide
welcome checks of our calculation.

For $|\Delta F| = 1$ processes the Wilson coefficients $C^\ren_m$ for the \sm{} can be found in \citeres{Bobeth:1999mk,Gorbahn:2004my} through \nnlo{}.
Thus, when neglecting penguin contributions, the ingredients for the flowed Wilson coefficients in \cref{eq:flowed-Wilson} are known through \nnlo{}.
For $|\Delta F| = 2$ processes the two operators in
\cref{eq:EW_physical_operators} become related by a Fierz identity and a suitable basis transformation is required.  In
this case, the \sm{} Wilson coefficient is known through
\nlo{}~\cite{Buchalla:1995vs}, with two contributions for Kaon mixing
known through \nnlo{}~\cite{Brod:2010mj,Brod:2011ty}.

\section{Conclusions}
\label{sec:conclusions}

In this contribution we constructed the flowed \ope{} for the
current-current operators of the electroweak Hamiltonian through \nnlo{}
in \qcd{}.  The non-renormalization of the flowed operators removes the
need to calculate the scheme transformation between the perturbative
Wilson coefficients and the matrix elements from lattice simulations
and, thus, could remove one source of uncertainties from theoretical
predictions of low-energy flavor observables.  Our complete results as
well as more details on our calculation and a comparison to the
\nlo\ results of of \citere{Suzuki:2020zue} will be presented elsewhere.

\acknowledgments

We thank the members of the CERN Theory Department, especially Martin L\"uscher, for initiating this project.
In addition, we thank him for providing us with his private notes.
Furthermore, we thank Ulrich Nierste and Benjamin Summ for stimulating discussions.
We acknowledge financial support by the \textit{Deutsche Forschungsgemeinschaft} (DFG, German Research Foundation) through grant \href{http://gepris.dfg.de/gepris/projekt/386986591?language=en}{386986591} and through the Collaborative Research Centre \href{http://p3h.particle.kit.edu/start}{TRR 257} funded through grant \href{http://gepris.dfg.de/gepris/projekt/396021762?language=en}{396021762}.

\appendix

\bibliographystyle{JHEP}
\bibliography{bib}

\end{document}